\documentclass[conference]{IEEEtran}
\usepackage[letterpaper,top=0.7in,bottom=1in,right=0.625in,left=0.625in]{geometry}
\usepackage{tikz}
\usetikzlibrary{shapes,arrows, positioning}
\IEEEoverridecommandlockouts
\usepackage{cite}
\usepackage{amsmath,amssymb,amsfonts}
\usepackage{algorithmic}
\usepackage{graphicx}
\usepackage{textcomp,multirow}
\usepackage{subcaption}
\usepackage{xcolor}
\usepackage{tikz}
\usepackage{soul}

\def\BibTeX{{\rm B\kern-.05em{\sc i\kern-.025em b}\kern-.08em
    T\kern-.1667em\lower.7ex\hbox{E}\kern-.125emX}}

\usepackage[colorinlistoftodos]{todonotes}

\begin{document}

\title{\LARGE{A Transformer-Based Framework for Payload Malware Detection and Classification}\\

}


\author{\normalsize Kyle Stein\(^1\), Arash Mahyari\(^{1,2}\), Guillermo Francia, III\(^3\), Eman El-Sheikh\(^3\)\\
\small \(^1\) Department of Intelligent Systems and Robotics, University of West Florida, Pensacola, FL, USA\\
\small \(^2\) Florida Institute For Human and Machine Cognition (IHMC), Pensacola, FL, USA\\
\small \(^3\) Center for Cybersecurity, University of West Florida, Pensacola, FL, USA\\
\small ks209@students.uwf.edu, amahyari@ihmc.org, gfranciaiii@uwf.edu, eelsheikh@uwf.edu}


\maketitle{}

\begin{abstract}
As malicious cyber threats become more sophisticated in breaching computer networks, the need for effective intrusion detection systems (IDSs) becomes crucial. Techniques such as Deep Packet Inspection (DPI) have been introduced to allow IDSs analyze the content of network packets, providing more context for identifying potential threats. IDSs traditionally rely on using anomaly-based and signature-based detection techniques to detect unrecognized and suspicious activity. Deep learning techniques have shown great potential in DPI for IDSs due to their efficiency in learning intricate patterns from the packet content being transmitted through the network. In this paper, we propose a revolutionary DPI algorithm based on transformers adapted for the purpose of detecting malicious traffic with a classifier head. Transformers learn the complex content of sequence data and generalize them well to similar scenarios thanks to their self-attention mechanism. Our proposed method uses the raw payload bytes that represent the packet contents and is deployed as man-in-the-middle. The payload bytes are used to detect malicious packets and classify their types. Experimental results on the UNSW-NB15 and CIC-IOT23 datasets demonstrate that our transformer-based model is effective in distinguishing malicious from benign traffic in the test dataset, attaining an average accuracy of 79\% using binary classification and 72\% on the multi-classification experiment, both using solely payload bytes. 
\end{abstract}


\section{Introduction}


As network traffic continues to evolve in modern society, the importance of Deep Packet Inspection (DPI) becomes an essential tool for the analysis of network packets and the security of networks. DPI goes beyond analyzing the network five-tuple, which consists of the source and destination IP address, source and destination port number, and the transport layer protocol, but also examines the data payload of each packet. The data payload of each packet consists of the content or information flowing through the network and processed by the respective end hosts. DPI dives into the actual content of data packets to extract valuable insights to identify threats or anomalies. The importance of DPI lies in the fact that adversaries can modify their MAC addresses or employ third-party devices, such as cloud services, to transmit malicious packets. These alterations make it challenging to identify such packets using only the traditional five-tuple approach. By thoroughly examining the payload of each packet, DPI can help distinguish between malicious or benign payloads, as well as the types of cyber attacks. Malicious payloads may include malware, viruses, or phishing attacks, while benign payloads include legitimate data and information exchanged between the network. Several research papers have contributed significantly to the field of DPI by proposing various approaches and methodologies.

In \cite{b2}, researchers describe an overview of the three different implementations of DPI. \textit{1) Signature-based Identification}: Relies on comparing the signatures (port numbers, string patterns, or bit sequences) of packets with known signatures in order to identify the associated application. Each application is associated with a specific signature that may include port numbers, string patterns, or bit sequences. By matching the signatures, DPI can recognize the data flow and determine the corresponding application. \textit{2) Application layer-based Identification}: Particularly useful for applications with distinct control and service features, this method zeroes in on the application gateway via the analysis of control and relative service flows. \textit{3) Behavior-based Identification}: Employed when data flows cannot be recognized by any known protocol. Instead, DPI analyzes user behavior or specific terminal characteristics to make judgments. For example, certain behaviors can be used to identify and filter out spam emails. The flow and relative service flow are examined to define the specific protocol associated with the behavior. Machine and deep learning models tend to fall under the category of behavior-based identification since they are designed to identify patterns and behaviors within the data.




Aceto \textit{et al.} \cite{b3} discuss the intersection of deep learning on mobile traffic classification. The authors discuss how deep learning models offer the potential to handle network traffic without relying on port information and can effectively distinguish between traffic generated by different applications. In \cite{Patheja}, authors employed various techniques, such as self-taught learning and soft-max regression. The study used the NSL-KDD as a benchmark dataset. Supervised learning techniques were used on over 41 statistical features, where none of the features were strict payload bytes. Doshi \textit{et al.} \cite{DDOS} leveraged machine learning for identifying Distributed Denial of Service (DDoS) attacks. Four machine learning models and one simple neural network were used to distinguish between normal Internet of Things (IoT) traffic and DDoS attacks. The paper focused on creating a novel dataset to test the algorithms and resulted in a large sample size of unbalanced malicious and benign samples.

 In \cite{Network}, machine learning techniques are leveraged to classify IP traffic on a 4G network. These researchers also generated their own dataset on the 4G network and applied common machine learning algorithms to the packet content of the dataset. The paper does not mention testing on malicious traffic, but rather the IP traffic being transmitted through the network. In \cite{IOT}, the authors applied machine and deep learning techniques to detect DDoS attacks on a sample dataset. None of the included features included raw payload bytes, but instead variables such as total number of packets, bytes, and total duration of packet transaction. 

DPI faces several current limitations that impact its effectiveness and performance. DPI is known to be resource-intensive since analyzing packet data requires substantial computational resources, especially since most networks have high volumes of traffic being transmitted through it. Thus, the state-of-the-art DPI algorithms limit the inspection of payloads to the initial bytes of packets. By only inspecting the initial portion of a packet's data, a system may overlook threats hidden deeper within the payload. Another challenge that DPI is currently facing is the growth of encrypted traffic being transmitted over networks. It is estimated by security researchers at Sophos that nearly 46\% of all malware in 2020 was hidden within an encrypted package \cite{sophos}. This challenge limits the ability to preserve privacy while inspecting payloads and requires parties to share their encryption keys with them to decrypt payloads, perform inspection and encrypt packet contents. 

The main contribution of this paper is the introduction of an algorithm for malware detection and classification based on transformers \cite{b1}. The proposed architecture capitalizes on the self-attention mechanism of transformers to capture the intricate patterns and dependencies present in the raw bytes of the network packet payloads, rather than relying on statistical-based features from packets. The payload contains the actual content of the network packet and would likely hold discernible patterns or signatures of malicious activity, while other information, such as the packet headers, are only meant for the transmission and management of data over a network and primarily contain addressing and protocol information. The proposed approach achieves not only enhanced accuracy in identifying malicious payloads, but also pushes the boundaries of current methodologies that help distinguish malicious payload types by employing two different classification heads. 

\section{Data}
\label{sec:data}
\vspace{-2mm}
\subsection{Datasets}

To evaluate the performance of the proposed method, we use several well-known and reputable datasets in this paper: the UNSW-NB15 \cite{UNSW} and CIC-IoT23 \cite{CICIOT}. The UNSW-NB15 dataset was created to overcome the shortcomings of the limited amount of publicly available intrusion detection network datasets, and includes 100 GB of raw network PCAP (Packet Capture) traffic, with various types of real and synthetic attacks. The CIC-IoT23 dataset aims to contribute a new and realistic IoT attack dataset, including seven different types of attacks categories. The data was recorded over an IoT topology composed of 105 devices, including raw network PCAP files for each type of attack and benign instances. 

\subsection{Data Pre-processing}
In this section, we describe the pre-processing steps to prepare the datasets. The UNSW-NB15 dataset offers an extensive set of ground truth labels and various components like the network five-tuple and attack timelines. From the available 100 GB of PCAP data, 4 GB were chosen to ensure efficiency throughout the study. For the CIC-IOT23 dataset, one benign and three attack PCAP files were chosen: Benign, Backdoor Malware, Vulnerability Attack, and Brute Force Attack.  

For our study, we are only interested in the TCP and the UDP transport layer information since these account for the majority of the network traffic on the transport layer. We discard packets that do not contain any payloads, corresponding to handshakes, acknowledgment, and any other network protocols and only focus on the packets that contain payloads. 

\subsubsection{UNSW-NB15}

Each PCAP file is processed by extracting the network five-tuple from each TCP or UDP packet, along with the timestamp and the corresponding transport layer payload. The transport layer payload bytes are converted to hexadecimal format, with all duplicate payload values omitted. The following step is to cross-reference the resulting dataset against the ground truth labels. By matching the rows based on IP addresses, ports, and adjusting the attack start and end time fields, we accurately label the benign and malicious network traffic flows. The network five-tuple is only used to cross-reference the ground truth labels, not as input into the model. The payload bytes are converted into hexadecimal and are then transformed into decimal integer format. These decimal integers are the primary input into the model architecture.

The outputs of the above process finalize the attack portion of the UNSW-NB15 dataset. The payload column data are selected and assigned labels of 1 for each row of malicious data. Similarly, we randomly select an equal number of benign payload entries as malicious entries to balance the final dataset. The benign entries are labeled as 0. It is important to randomly select an equal amount of benign and malicious entries to ensure the model is not biased toward the majority class, which could lead to a significant number of false positives or negatives. The finalized binary classification dataset contains an equal amount of benign and malicious entries.

For the multi-class dataset, we specifically analyze three types of attacks: Fuzzers, Exploits, and Generic. These attacks were selected because they are known for their complex and sophisticated payloads, in contrast to attacks like distributed denial of service (DDoS), which primarily rely on overwhelming volume rather than payload complexity. Fuzzers aim to disrupt the functioning of a program or network by feeding it randomly generated data. Exploits are attacks that take advantage of known security vulnerabilities within an operating system or a piece of software. Generic refers to a broad category of attacks that are effective against all block ciphers, regardless of the block or key size, without considering the specific structure of the block-cipher. This could involve a variety of methods to undermine encryption and data integrity.

\subsubsection{CIC-IOT23}

The pre-processing steps for the CIC-IOT23 dataset are similar to that of the UNSW-NB15. The processing starts with the extraction of the respective transport layer payload from every TCP or UDP packet for each of the four selected PCAP files. These payload bytes are then transformed into hexadecimal format, ensuring once again the removal of any duplicate payloads. Each extracted sample of attack payload bytes are concatenated to form one dataset, while other resulting dataset consists of benign payload bytes. 

Following these steps, the attack and benign datasets are established. The payload bytes are converted to hexadecimal format and are then converted to decimal integers. Every row of malicious data is tagged with a label of 1. To ensure balance in the dataset and prevent biases in the model that could potentially lead to a high number of false positives or negatives, an equal number of benign payload entries are randomly chosen to match the malicious ones. All benign traffic entries are consequently labeled as 0 for the classification analysis.

Similarly to the UNSW-NB15 dataset, our study specifically focuses on three types of attacks for the CIC-IOT23: Backdoor Malware, Vulnerability Attack, and Brute Force Attack. These attacks rely on the specific bytes within the payload to carry out their malicious goals, similar to the attacks chosen from the UNSW-NB15 dataset. Backdoor Malware attacks demonstrate how unauthorized remote access can be gained by attackers. Vulnerability Attacks aim to show the exploitation of weaknesses within the IoT devices or network. Brute force attacks are attempts to crack passwords or encryption through trial and error.

\section{Architecture and Model Training}
\label{sec:arch}

In this section, we describe the architecture, training, and evaluation of our model for deep packet inspection.

\subsection{Architecture}

\begin{figure*}[t]
\centering
\includegraphics[width=0.7\linewidth]{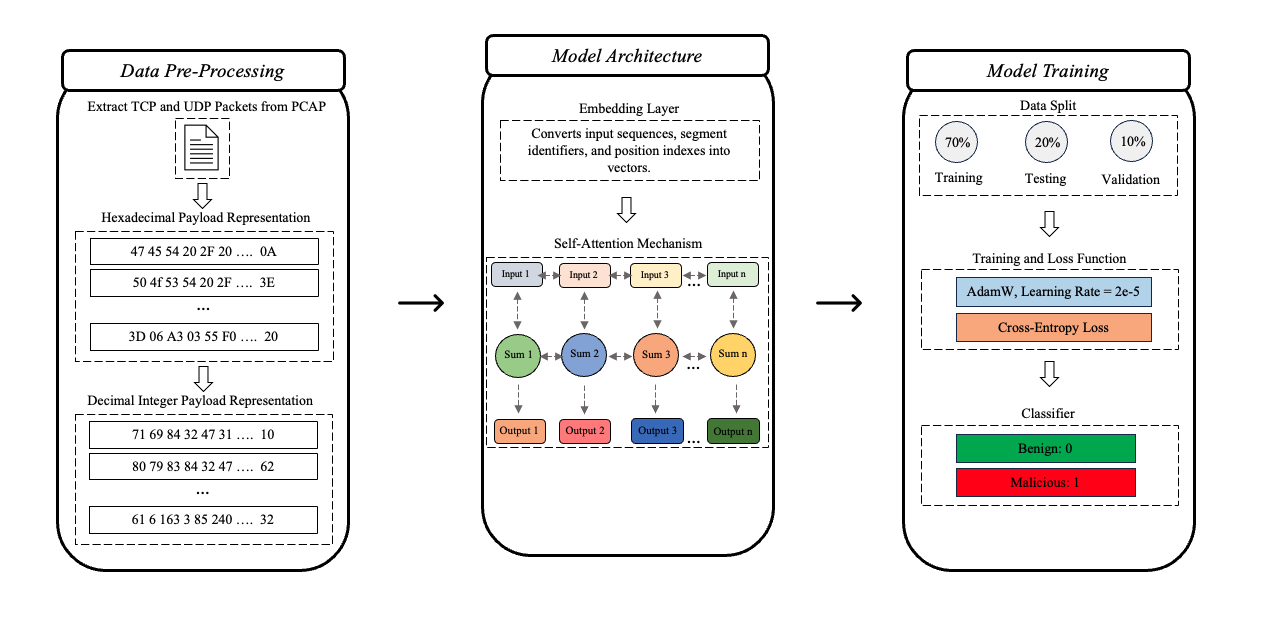}
\caption{\textbf{The overall architecture of the proposed packet detection algorithm. Represented here is the payload as input for the model. First, the input payload is pre-processed and converted to hexadecimal format. Next, every two hexadecimal characters are converted to a decimal integer value between [0, 255]. The integer strings are then fed into the embedding layer to obtain its embedding vector, then processed in the self attention mechanism. The classifier head then predicts if the packet is either benign or malicious.}}
\label{fig:architecture}
\end{figure*}

Transformer-based models have been powerful in natural language processing tasks due to their ability to understand the relationships between words in sequential textual data. Although individual bytes in a network packet may not inherently carry semantic meanings like words in a sentence, their sequences, patterns, and relative positions can encapsulate information about the nature of the packet. Leveraging transformer's capability to capture contextual patterns, we show that it can be effectively applied to discern distinctions in packet sequences. This process is powered by the self-attention mechanism \cite{Attention}, enabling each byte in the input sequence to reference and weigh other bytes within that sequence while formulating its output representation. This determines a significance score for each byte in the sequence, denoting the degree of attention it should receive.

The model architecture can be divided into three main parts:
\begin{itemize}
\item \textbf{Embedding Layer:} The initial layer converts input sequences, segment identifiers, and position indexes into vectors. These input sequences are derived from the encoded and padded data prepared during the preprocessing phase. 

\item \textbf{Transformer Blocks and Self-Attention Mechanism:} The transformer blocks form the core of the model, with each block comprised of a transformer encoder that integrates a multi-head self-attention mechanism and a position-wise fully connected feed-forward network \cite{b6}. Within our model are 12 of these transformer blocks.  The self-attention mechanism is what sets transformers apart from other traditional deep learning architectures. The self-attention mechanism can attend to all positions in the sequence concurrently, which is necessary in capturing long-range interdependencies. With the aid of positional encodings, transformers can maintain the sequence's order, an essential feature for sequential data such as network packets.

\item \textbf{Output Layer:} The final hidden state corresponding to the first token in each sequence is used as the aggregate sequence representation for classification tasks. In our case, it is used to determine whether a network packet is benign or malicious.
\end{itemize}

The selected hyperparameters of the model include the number of unique bytes (\textit{i.e.} 256), hidden size, number of hidden layers, number of attention heads, the intermediate size, maximum length of the input payload,  and the number of labels for this application, and are presented in Table~\ref{tab:bert_config}.

\begin{table}[h] \label{tab:bert}
\vspace{-2mm}
\centering
\caption{Model Configurations}
\label{tab:bert_config}
\begin{tabular}{|c|c|}
\hline
\textbf{Configuration Parameter} & \textbf{Value} \\
\hline
unique\_bytes & 256 \\
\hline
hidden\_size & 768 \\
\hline
num\_hidden\_layers & 12 \\
\hline
num\_attention\_heads & 12 \\
\hline
intermediate\_size & 3072 \\
\hline
max\_position\_embeddings & 1460 \\
\hline
num\_labels & 2 or 3\\
\hline
\end{tabular}

\vspace{-2mm}
\end{table}

The unique bytes parameter represents the total number of unique elements found in the input data, totalling 256, for each unique hexadecimal value. The hidden size parameter is essential in determining the dimension of the hidden state. The number of hidden layers represents the number of stacked transformer layers present in the model. Complementing these layers is the number of attention heads, which control the model's ability to attend to different segments of the input. This is pivotal for the model to comprehend and learn from the input data effectively.

The intermediate size determines the dimension of the intermediate later in the network within each transformer layer. This value helps ensure that the network can process and transform the input data at each layer effectively. The maximum position embeddings denote the maximum length of the input sequence. For this study, we do not restrict the maximum payload length, which is up to 1460 bytes. The last hyperparameter is the number of labels, which is set to 2 for binary classification or 3 for multi-class classification. These configurations remained consistent through our study of analyzing the payload bytes.

\subsection{Model Training}

The model is trained using a 70-20-10 split: 70\% of the data is used for training, 20\% is used for testing, and 10\% is used for validation. To enhance the model's generalization across all classes, each class is balanced to have the same number of samples, ensuring fair representation of each class throughout model training. The training process of the transformer and classification head is conducted end-to-end. The transformer is responsible for encoding the input sequences into representations that capture the patterns and dependencies in the data. The classification head takes these representations and maps them to respective class labels.  

Cross-entropy loss is used as the cost function to train the model and the AdamW optimizer is used with a learning rate of 2e-5. The AdamW optimizer provides weight decay regularization, an approach which is crucial as it limits the magnitude of the weights, preventing the model from becoming overly complex and generalized to the training data, in-turn resulting in the mitigation of overfitting \cite{b7}. 

The model is trained for 5 epochs and a scheduler for learning rate decay is also used to reduce the learning rate over the training period. The scheduler gradually increases the learning rate from zero to a specified learning rate during the warmup period, then linearly decreases the learning rate over the remaining epochs of training \cite{b8}. This process encourages the model to find a more generalized solution for the test and validation sets, rather than just the training set. This learning rate strategy also benefits in the prevention of overfitting, as it prevents the model from converging too quickly to a solution that may be specific to the training data. The training of the model was conducted on NVIDIA GeForce RTX 2080 GPUs. This state-of-the-art hardware enabled us to harness significant computational power, facilitating faster processing and more efficient learning from the datasets. 

\begin{table*}[t!]
\centering
\caption{Results of Binary Classification of Malware Detection on the Test Dataset}
\label{tab:binary}
\begin{tabular}{|l|l|l|l|l|l|l|l|l|}
\hline
{\multirow{2}{*}{\textbf{Method}}} & \multicolumn{4}{c|}{\textbf{UNSW-NB15}} & \multicolumn{4}{c|}{\textbf{CIC-IOT23}}\\
\cline{2-9}
& Accuracy & Precision & Recall & F1-Score & Accuracy & Precision & Recall & F1-Score \\
\hline

1D-CNN \cite{b9} & 74.98 & 68.37 & 93.06 & 78.82 & 75.30 & 72.63 & 81.72 & 76.91 \\ \hline

2D-CNN \cite{b10} & 75.56 & 68.41 & \textbf{95.46} & 75.56 & 72.19 & 68.47 & 82.54 & 74.85  \\ \hline

LSTM \cite{b10} & 71.65 & 69.33 & 77.71 & 73.28 & 71.60 & 71.18 & 72.65 & 71.91  \\ \hline

\textbf{Proposed Method} & \textbf{79.57} & \textbf{73.26} & 93.16 & \textbf{79.57} & \textbf{79.07} & \textbf{73.79} & \textbf{90.38} & \textbf{81.25}\\ \hline
\end{tabular}

\label{table:performance_metrics}
\end{table*}

\begin{table*}[t!]
\centering
\caption{Results of Multiclass Classification of Malware Types on the Test Dataset}
\label{tab:multi}
\begin{tabular}{|l|l|l|l|l|l|l|l|l|}
\hline
{\multirow{2}{*}{\textbf{Method}}} & \multicolumn{4}{c|}{\textbf{UNSW-NB15}} & \multicolumn{4}{c|}{\textbf{CIC-IOT23}}\\
\cline{2-9}
& Accuracy & Precision & Recall & F1-Score & Accuracy & Precision & Recall & F1-Score \\
\hline

1D-CNN \cite{b9} & 71.60 & 72.50 & 71.60 & 71.93 & 62.61 & 66.75 & 62.61 & 62.45 \\ \hline

2D-CNN \cite{b10} & 72.41 & \textbf{77.17} & 72.41 & 73.22 & 61.05 & 66.28 & 61.05 & 60.60 \\ \hline

LSTM \cite{b10} & 69.98 & 70.21 & 69.83 & 69.92 & 60.39 & 62.94 & 60.41 & 60.24 \\ \hline

\textbf{Proposed Method} & \textbf{74.24} & 76.34 & \textbf{74.24} & \textbf{74.61} & \textbf{69.25} & \textbf{70.51} & \textbf{69.25} & \textbf{69.31} \\ \hline
\end{tabular}

\label{table:performance_metrics}

\end{table*}

\section{Encrypted Traffic}
\label{sec:threat}
In our study, we focus on the raw payload bytes within TCP and UDP network packets from the discussed datasets. However, it is crucial to note the limitations when dealing with encrypted traffic. Cryptography--the process of encrypting the data--secures the transmission of data flowing through a network, making the data unreadable to unauthorized users without the proper keys. When the plaintext is encrypted with a key resulting in the ciphertext, the ciphertext gives no information about the plaintext \cite{menezes2018handbook}. Let's assume $m$ is the random variable representing the raw payload message and $E_k(.)$ is the encryption algorithm that encrypts the message with the key $k$. Then, knowing the ciphertext $c_1$ reveals absolutely no information about the plaintext $m_1$: $p(m=m_1|E_k(m1)=c_1)=p(m=m_1)$. In plain words, every time the plaintext $m_1$ is decrypted with the same cryptography algorithm $E_k(.)$ and the same encryption key $k$, it results in a different ciphertext, thus revealing no information about the plaintext \cite{menezes2018handbook}. While several studies have been able to classify the traffic of different applications from the cipher text \cite{android} \cite{socialmedia}, their success is attributed to the fact that different applications use different random generators for encryptions which appears as a signature in their encrypted payloads but still reveal no information about the plaintext \cite{deeppacket}. On the other hand, malware detection algorithms (including the proposed algorithm in this paper) requires access to the information of the plaintext that cannot be revealed by the ciphertext. If any algorithm is able to detect the signature of malware from the ciphertext, it means that the encryption algorithm is not strong enough to hide the information about the plaintext which means $p(m=m_1|E_k(m1)=c_1) \neq p(m=m_1)$. We have conducted several experiments to demonstrate this.

We implemented both Advanced Encryption Standard (AES) encryption \cite{b11} and Fernet symmetric encryption \cite{fernet} on the payloads. AES takes the raw payload data, a 256-bit key, and a 16-byte initialization vector (IV) to produce the encrypted output. Each key and IV combination ensures the uniqueness of the encryption process. Fernet uses AES in Cipher Block Chaining (CBC) mode with a 128-bit key for encryption. After encrypting the payloads, similar procedures discussed in the methodology above were implemented. Each encrypted payload, in hexadecimal format, was converted to a decimal integer sequence and then padded to ensure uniform length for each input.

These inputs were then trained and tested in a similar manner as above with the same architecture. The AES cryptologic algorithm results showed that the model was not able to effectively distinguish between malicious and benign payloads, with a test accuracy of 57.16\% and an F1-Score of 44.52\%. This demonstrates that encryption algorithms like AES successfully diminish the learnable patterns in the raw bytes, which our model relies on to classify the payloads. However, the Fernet algorithm produced a test accuracy of 91.41\% and an F1-Score of 92.09\%. This experiment shows that certain encryption algorithms may not be strong enough to hide the plaintext information, while others can.

\section{Results}
\label{sec:results}

To evaluate the performance of the proposed method, several metrics are taken into consideration. For this study, we consider accuracy, precision, recall, and F1-Score on the test dataset. It is important to note that the test dataset used in our evaluation process differs from the training dataset. The test dataset consists of randomly selected, unseen samples and ensures that the model is using a distinct dataset to evaluate the performance. 








The proposed method is compared against other state-of-the-art models \cite{b9} \cite{b10}. Our primary intention behind comparing a transformer model with these deep learning algorithms lies in evaluating different sequence processing architectures. While all of these models fundamentally process sequences, their mechanisms are distinct. Convolutional Neural Networks (CNNs) capture localized patterns and hierarchical structures in data. This is valuable when analyzing patterns emerging from chunks of network packets. Long Short-Term Memory networks (LSTMs) utilize recurrent connections to remember patterns over long sequences. However, our transformer-based model is designed to comprehend and capture context over a wider range without the recursive nature of LSTMs. The self-attention mechanism allows the model to weigh significance of different parts of the sequence, providing a global understanding of data.

Our study differs in several aspects. We first eliminated duplicate payload values, which allowed the model to analyze unique payloads, increasing the data quality to better generalize to unseen data. Additionally, our input was not limited to the inital 784 payload bytes as seen in \cite{b9}, but used the maximum of 1460 bytes. The typical Maximum Transmission Unit (MTU) which can be sent through a packet-based network is 1500 bytes, minus 40 bytes for the header information. The header data was omitted to reduce bias that may be introduced by the networking addresses and protocol details. By utilizing strict payload bytes, we ensured our model had access to all available information within each payload, which is crucial since malicious data may be hidden deeper into the payload of some packets. The limitation of discarding portions of byte data is addressed, showing a more global nature of the complexity of each payload. The evaluation results for the classifiers can be found in Tables \ref{tab:binary} and \ref{tab:multi}. We assessed the performance of these classifiers under two conditions: binary classification and multi-class classification tasks.

\subsection{Binary Classification}
Table II displays the performance comparisons of the stateof-the-art methods for malware detection in a binary classification setting. Across the two datasets used for evaluation, the proposed method consistently outperformed the other techniques. For the UNSW-NB15 dataset, the proposed method achieved the highest accuracy of 79.57\%, superior to 1D-CNN, 2D-CNN, and LSTM, which scored 74.98\%, 75.56\%, and 71.65\%, respectively. The great performance of the proposed method is attributed to its ability to capture the context using its self-attention mechanism. In terms of F1-Score, the proposed method reported the best score at 79.57\%, while 1D-CNN, 2D-CNN, and LSTM secured 78.82\%, 75.56\%, and 73.28\%, respectively. However, the recall of the proposed method was slightly less compared to the 2D-CNN. This indicates a potential area of improvement in minimizing our method’s false negatives, which is crucial for reducing the risk of undetected malware.

For the CIC-IOT23 dataset, our method outperformed several benchmarks. Our proposed method achieved an accuracy of 79.07\% and an F1-Score of 81.25\%. In comparison, the 1DCNN yielded an accuracy of 75.30\% with an F1-Score of 76.91\%, the 2D-CNN had an accuracy of 72.19\% and an F1-Score of 74.85\%, and the LSTM delivered an accuracy of 71.60\% alongside an F1-Score of 71.91\%. The results indicate that the proposed method, when trained on either of the datasets, provides a significant performance improvement over the compared models across the majority of metrics. This demonstrates the valuable information that the payload carries for network intrusion detection tasks.

\subsection{Multi-Class Classification}
Table III reveals significant improvements in multi-class malware detection compared to state-of-the-art methods on both datasets. Each dataset consisted of three different types of malware, referenced above in the data pre-processing section. For the UNSW-NB15 dataset, the proposed method achieved the highest performance metrics across most categories: an accuracy of 74.24\%, recall of 74.24\%, and F1-Score of 74.61\%. In comparison, the 2D-CNN method achieved an accuracy of 72.41\%, the highest precision of 77.17\%, recall of 72.41\%, and F1-Score of 73.22\%. The 1D-CNN method resulted in an accuracy of 71.60\%, precision of 72.50\%, recall of 71.60\%, and F1-Score of 71.93\%. The LSTM method produced an accuracy of 69.98\%, precision of 70.21\%, recall of 69.83\%, and F1-Score of 69.92\%.
When applied to the CIC-IOT23 dataset, the proposed method again outperformed the comparative methods with an accuracy of 69.25\%, precision of 70.51\%, recall of 69.25\%, and F1-Score of 69.31\%. The 1D-CNN achieved metrics of 62.61\% in both accuracy and recall, 66.75\% in precision, and 62.45\% in F1-Score. The 2D-CNN metrics showed 61.05\% for accuracy and recall, 66.28\% for precision, and 60.60\% for F1-Score. The LSTM model’s performance once again produces the lowest results at 60.39\% for accuracy, 62.94\% for precision, 60.41\% for recall, and 60.24\% for F1-Score. The findings suggest that the proposed method for multi-class classification demonstrates enhanced performance in identifying the three distinct types of attacks within each dataset, compared to the previous methodologies.  This further highlights the impact our proposed approach has in bolstering network intrusion detection, reinforcing the performance of our methodology for cybersecurity defenses.

\section{Conclusion}
\label{sec:conclusion}
This paper explored the application of a transformer-based model for malware detection and classification. The model was evaluated on the UNSW-NB15 and CIC-IOT23 datasets, focusing on the payloads of UDP and TCP packets serving as inputs. Our method produced robust results on the classification of benign versus malicious packets when compared to state-of-the-art methods. Using the payload bytes as input vectors, we were able to classify when different packets are either benign or malicious, as well as which types of attacks were present.  While the payload bytes of packets are significantly different from the natural language structure, this study shows that the transformer-based model developed for natural language can be leveraged to capture and learn the intricate sequential patterns of the payload bytes.


\section{Acknowledgement}
\label{sec:ack}
This work is partially supported by the UWF Argo Cyber Emerging Scholars (ACES) program funded by the National Science Foundation (NSF) CyberCorps® Scholarship for Service (SFS) award under grant number 1946442. Any opinions, findings, and conclusions or recommendations expressed in this document are those of the authors and do not necessarily reflect the views of the NSF.

\vspace{12pt}

\end{document}